\documentclass[letterpaper,10pt]{article}
\usepackage{tabularx} % extra features for tabular environment
\usepackage{amsmath}  % improve math presentation
\usepackage{graphicx} % takes care of graphic including machinery
\usepackage[margin=1in,letterpaper]{geometry} % decreases margins
\usepackage{cite} % takes care of citations
\usepackage[final]{hyperref} % adds hyper links inside the generated pdf file
\usepackage{authblk}
\usepackage{xspace}
\usepackage{algorithm2e}
\usepackage{multirow}
\usepackage{floatrow}
\usepackage{cleveref}
\usepackage{comment}
\usepackage{enumitem}
\usepackage[dvipsnames]{xcolor}
\usepackage{diagbox}
\hypersetup{
	colorlinks=true,       % false: boxed links; true: colored links
	linkcolor=blue,        % color of internal links
	citecolor=blue,        % color of links to bibliography
	filecolor=magenta,     % color of file links
	urlcolor=blue         
}
\begin{document}

\title{Semi-supervised Graph Neural Networks for Pileup Noise Removal}
\author[1]{Tianchun Li$^*$}
\author[1]{Shikun Liu$^*$}
\author[2]{Yongbin Feng$^*$}
\author[1]{Garyfallia Paspalaki}
\author[2]{Nhan V. Tran}
\author[1]{Miaoyuan Liu}
\author[1]{Pan Li}
\affil[1]{Purdue University, West Lafayette, IN 47907}
\affil[2]{Fermi National Accelerator Laboratory, Batavia, IL 60510}
\date{}                     %% if you don't need date to appear
\setcounter{Maxaffil}{0}
\renewcommand\Affilfont{\itshape\small}

\newcommand{\pan}[1]{{\color{blue}{P: #1}}}
\newcommand{\tianchun}[1]{{\color{ForestGreen}{T: #1}}}
\newcommand{\shikun}[1]{{\color{magenta}{S: #1}}}
\newcommand{\yongbin}[1]{{\color{purple}{YB: #1}}}
\newcommand{\nhan}[1]{{\color{cyan}{NT: #1}}}
\newcommand{\mia}[1]{{\color{orange}{Mia: #1}}}
\newcommand{\lisa}[1]{{\color{teal}{lisa: #1}}}
\newcommand{\todo}[1]{{\color{red}{ToDos: #1}}}
\newcommand{\pt}{\ensuremath{p_\mathrm{T}}\xspace}
\newcommand{\GeV}{\,\ensuremath{\mathrm{GeV}}\xspace}
\newcommand{\Lagr}{\mathcal{L}}
\newcommand\nnfootnote[1]{%
  \begin{NoHyper}
  \renewcommand\thefootnote{*}\footnote{#1}%
  \addtocounter{footnote}{-1}%
  \end{NoHyper}
}

\date{\today}
\maketitle

\def\thefootnote{*}\footnotetext{These authors contributed equally to this work}
% \nnfootnote{These authors contributed equally to this work}

\begin{abstract}
The high instantaneous luminosity of the CERN Large Hadron Collider leads to multiple proton-proton interactions in the same or nearby bunch crossings (pileup). Advanced pileup mitigation algorithms are designed to remove this noise from pileup particles and improve the performance of crucial physics observables. This study implements a semi-supervised graph neural network for particle-level pileup noise removal, by identifying individual particles produced from pileup. The graph neural network is firstly trained on charged particles with known labels, which can be obtained from detector measurements on data or simulation, and then inferred on neutral particles for which such labels are missing. This semi-supervised approach does not depend on the ground truth information from simulation and thus allows us to perform training directly on experimental data. The performance of this approach is found to be consistently better than widely-used domain algorithms and comparable to the fully-supervised training using simulation truth information. The study serves as the first attempt at applying semi-supervised learning techniques to pileup mitigation, and opens up a new direction of fully data-driven machine learning pileup mitigation studies.

\end{abstract}

\section{Introduction}\label{chp:intro}
The high instantaneous luminosity of the CERN Large Hadron Collider (LHC) enables studies of the deep mysteries of our universe, such as the nature of the Higgs boson~\cite{ATLAS:2012yve,CMS:2012qbp} and dark matter as well as the origin of the matter-antimatter asymmetry~\cite{HEPAPSubcommittee:2014bsm}. The enormous amount of data coming from increasingly noisy particle collisions, recorded by more complex detectors, poses various challenges to data collection and analysis~\cite{ATLAS:2020cli,ATLAS:2017ywy,ATLAS:2018txj,Sirunyan:2020foa,CMS:2019ctu}. Multiple collisions in the same or nearby proton bunch crossings lead to overlapping particle interactions, referred to as pileup (PU). To achieve the desired physics sensitivity with the LHC data, the noise from PU particles needs to be identified and mitigated effectively in order to identify signals of interest, i.e., those from the primary interaction of interest, often referred to as the leading vertex (LV). The average number of PU interactions during the LHC data-taking period of 2016 to 2018 is around 30-40~\cite{ATLAS:2019pzw,CMS:2021xjt}. This is expected to increase in future data-taking periods and reaches around 150 for the high luminosity LHC~\cite{Bruning:2015dfu}. Improvements in pileup mitigation techniques can therefore have significant effects on the entire current and future LHC program, through performance gains in the reconstruction of all high-level physics objects which in turn are used in nearly all measurements and searches at ATLAS and CMS.

Particles produced from proton-proton (pp) interactions are reconstructed using the hit information in the tracking detectors and the energy deposits in the calorimeters. Due to the excellent performance of the charged particle tracking systems and their reconstruction algorithms, the track and vertex information of charged particles within the tracker acceptance can be precisely determined~\cite{ATLAS:2017kyn,CMS:2014pgm}. Most charged particles associated with PU vertices can be identified and removed from the event. This is often referred to as charged hadron subtraction, and its performance can be found in~\cite{ATLAS:2020cli,Sirunyan:2017ulk}. The remaining challenge of the pileup mitigation task falls therefore mostly on neutral particles, including photons and neutral hadrons.

During data-taking between 2009-2012, most of the developments of pileup mitigation algorithms focused on area-based subtractions~\cite{Cacciari:2008gd,Krohn:2013lba,Berta:2014eza,Cacciari:2014jta}, which correct the physics quantities based on the average pileup density per event. While these methods provide unbiased estimations of jet four-momenta, their resolutions usually become worse and only operate at the level of a whole jet object. More advanced particle-level algorithms have been developed later on, such as SoftKiller~\cite{Cacciari:2014gra} and PUPPI~\cite{Bertolini:2014bba}. SoftKiller makes use of the fact that particles from PU vertices tend to have lower transverse momentum (\pt) than the particles from the LV, and applies a \pt cut to remove low-\pt (``soft") pileup particles. PUPPI, on the other hand, makes use of the neighboring particle information and defines a local shape variable $\alpha$ for each particle. Per-particle weights are calculated based on $\alpha$, and the particle four momenta are rescaled with their corresponding weights. PUPPI has achieved significantly better performance compared with other methods and has been adopted in many LHC analyses~\cite{Sirunyan:2020foa}. All of these rule-based algorithms developed and described above do not need labeled simulated data for training, but the parameters in these algorithms need to be carefully tuned based on the real data for each experimental setting.

With the recent rapid developments of machine learning (ML) algorithms, studies~\cite{Komiske:2017ubm,Martinez:2018fwc,Mikuni:2020wpr,Maier:2021ymx} have been performed applying ML techniques to the pileup mitigation task. These ML-based algorithms adopt convolutional neural network~\cite{Komiske:2017ubm}, gated graph neural network~\cite{li2017gated,Martinez:2018fwc}, and attention-based models~\cite{Mikuni:2020wpr,Maier:2021ymx}, to learn complex patterns from the training data and have achieved significantly better performance than the classical domain algorithms in simulation studies. Most of these algorithms require a large amount of LV/PU label information of input particles to get sufficiently trained, termed ``fully-supervised'' methods. However, such label information for neutral particles is intractable to retrieve in the full Geant-based simulations~\cite{Agostinelli:2002hh}, and does not exist in real collision data. The simulation inaccuracy makes it non-trivial to train and deploy these algorithms to the actual experiments. Dedicated model tunings and precise calibrations are often required, bringing in extra work and systematic uncertainties~\cite{Sirunyan:2020foa,CMS:2020poo,ATLAS:2018xcf}.

The goal of our work is to abandon the previous fully-supervised methods as they rely on the label information of neutral particles. Instead, a novel semi-supervised machine-learning technique (SSL) is applied, taking advantage of the fact that the LV/PU labels of charged particles can still be precisely determined with reconstruction-level information of real collision data. Inspired by the success of PUPPI, our key idea is to capture the effects of neighboring particle features on the LV/PU estimation of the target particle, which does not strongly depend on whether the target particle is neutral or charged. To achieve this, we first construct a graph connecting particles close to each other in the physical space, and then train a graph neural network (GNN) using exclusively the LV/PU labels of charged particles. The trained GNN is further applied to neutral particles to estimate the probability of each of them being produced from LV or PU. To avoid the label leakage and the potential bias due to the feature shifts from charged to neutral particles, we propose a random masking technique, which can be viewed as a separate and unique contribution to the adopted machine learning technique itself. The GNN mimics PUPPI in the sense that it explores the neighboring particle features to form a \emph{data-driven} local shape variable for pileup mitigation, fully learned from the real experimental data. It not only aggregates the features in a more expressive way than PUPPI, but also avoids the complex manually tuning procedure of PUPPI.

The effectiveness of this SSL approach is carefully studied and confirmed by comparing the performance of a GNN from fully-supervised training, a GNN with the same architecture but from semi-supervised training, and the domain algorithm PUPPI, in the simulations of different processes and different pileup conditions. {\tt DELPHES}-based~\cite{deFavereau:2013fsa} simulation samples are used in order to carry out the fully-supervised training, with more details provided in later sections. It has been found that there is no significant performance drop going from fully supervised training to semi-supervised training, and the GNNs achieve better performance than PUPPI in both cases .

The studies in this paper serve as the first attempt of applying a SSL approach to pileup mitigation studies. This approach does not rely on any ground truth information from simulations, and therefore the full workflow can be performed directly on real collision data, without concerns about differences between data and simulation or imperfect choices in the truth labeling. Comparisons are made between techniques using simulated data with truth labels. With promising results, it is worthwhile to study and explore similar approaches in more realistic simulations and real collision data in the near future.

Details of our studies and the results are presented in the following sections. Section~\ref{chp:related} provides a brief overview of the previous related works. Section~\ref{chp:dataset} describes the details of the simulation setup and the dataset used. Section~\ref{chp:method} presents the methodology of the semi-supervised training technique, the network architecture, and the training setup. Section~\ref{chp:results} presents the results, with the performance benchmarks of labeling LV/PU particles, and the subsequently reconstructed physics quantities such as observables of hadronic jets and missing transverse momentum. Section~\ref{chp:discussion} discusses the results and followup studies. Section~\ref{chp:summary} summaries the paper with an outlook for future developments.
\section{Related work}\label{chp:related}
As briefly introduced in Section~\ref{chp:intro}, SoftKiller and PUPPI are the two currently widely used pileup mitigation algorithms which operate on a per-particle basis. SoftKiller breaks one event into patches and defines a single $\pt$ cut $\pt^{\mathrm{cut}}$ based on the \pt of the hardest particle in each patch $p^{\mathrm{max}}_{\mathrm{T},i}$: 
\begin{equation}
    \pt^{\mathrm{cut}} = \mathrm{median}_{i\in\mathrm{patches}}\{p^{\mathrm{max}}_{\mathrm{T},i}\}
\end{equation}
Particles with \pt lower than $\pt^{\mathrm{cut}}$ will be marked as pileup and removed from the event in the subsequent reconstructions. Compared with previous area-based pileup mitigation algorithms~\cite{Cacciari:2008gd,Krohn:2013lba,Berta:2014eza,Cacciari:2014jta}, SoftKiller operates at the individual particle level and brings significant improvements to the hadronic jet observables, such as mass, \pt, and substructure variables. However, on the other hand, making use of only the \pt information will drop lots of other useful information.

PUPPI makes use of the information from local neighboring particles. The local shape variable $\alpha$ is calculated according to:
\begin{equation} \label{eq:local-shapre}
    \alpha_i = \log\sum_{j}\xi_{ij}\times\theta(R_\mathrm{min}<\Delta R_{ij}<R_{0})
\end{equation}
where $\Delta R_{ij}=\sqrt{(\Delta\eta_{ij})^2 + (\Delta\phi_{ij})^2}$ is the distance between the neighboring particle $j$ and the target particle $i$ in the $\eta-\phi$ space; $\eta$ is pseudorapidity and $\phi$ is the azimuthal angle in collider cylindrical space; the sum $j$ is over neighboring particles in the event with $\Delta R<R_0$ and $\Delta R>R_\mathrm{min}$; $\xi_{ij}=p_{\mathrm{T},j}/\Delta R_{ij}$; and $\theta$ is the Heaviside step function. The local shape $\alpha$ is computed per particle, and PUPPI weights are assigned to individual particles accordingly, which describes the probability of each particle being produced from LV. The particle four momenta are rescaled based on the PUPPI weights. 

Compared with SoftKiller, PUPPI makes better use of the local neighboring features and the $\alpha$ calculation does not depend on target particles' \pt. But the critical part of the PUPPI algorithm is one \textit{ad hoc}, expert-level metric. There are some parameters that require extensive studies and manual tunings, such as the choice of the cone size $R_0$, the selection of the neighboring particles, and the metric $\xi_{ij}$, which can be sometimes changed to $(p_{\mathrm{T},j}/\Delta R_{ij})^2$, etc. Hence, more recent efforts have focused on developing machine learning approaches to automatically learn such combinations.

PUMML serves as the first attempt to apply modern deep learning (DL)
techniques for pileup mitigation. It treats collision events as images and particles as pixels in the $\eta-\phi$ grids. With a convolutional neural network applied to extract local features, it achieves better performance compared with PUPPI and SoftKiller. However, representing particles as images requires fixed spatial resolution, which in realistic cases depends on the $\eta-\phi$ positions and can vary dramatically for different types of particles. For sparse events with a limited number of particles in certain regions, flat images would also waste computing resources.

Benefiting from the rapid developments in the DL community, graph neural networks and attention mechanisms are introduced in particle physics~\cite{Shlomi:2020gdn}, such as jet flavor tagging~\cite{Komiske:2018cqr,Qu:2019gqs}, calorimeter and event reconstruction~\cite{Qasim:2019otl,Ju:2020xty,Pata:2021oez}, and also pileup mitigation studies~\cite{Martinez:2018fwc,Mikuni:2020wpr,Maier:2021ymx}. Treating each particle as one unit, these DL architectures do not assume regular detector geometry and can explore much more effectively and efficiently the local structures in collision events. Models including PUPPIML~\cite{Martinez:2018fwc}, ABCNet~\cite{Mikuni:2020wpr}, and PUMA~\cite{Maier:2021ymx} belong to this category and have shown promising results produced on {\tt DELPHES}-based~\cite{deFavereau:2013fsa} simulation data.

Applying these similar architectures on more realistic {\tt GEANT}-based~\cite{Agostinelli:2002hh} simulations and real collision data is the next major task. However, it is very challenging to apply such ML algorithms to these more realistic scenarios as the proposed models need full supervision, i.e., being trained with a large number of labeled (LV or PU) neutral particles. The ground-truth label information is hard to be recovered in {\tt GEANT} simulations and does not exist in real collision data. In order to overcome this challenge and bring these powerful DL models into the realistic deployment and usage for the LHC experiments, we explore the idea of semi-supervised learning, where the training is performed on charged particles, whose LV/PU labels can be determined at reconstruction level for both data and simulations, and the trained model is then applied on neutral particles to estimate their LV/PU probabilities.
\section{Datasets}\label{chp:dataset}
For our studies, simulated datasets have been generated of different physical processes under different pileup conditions. In this study, we select three pileup conditions where the numbers of pileup interactions (n$_{\text{PU}}$) are 20, 80, and 140 respectively, and two hard scattering signal processes, $\mathrm{Z}(\nu\nu)+$jets and $\mathrm{H}(b\bar{b})+$jets. 
We study these two signal processes because they include important physics signatures which are affected significantly by additional pileup interactions.  In the $\mathrm{Z}(\nu\nu)+$jets process, the invisibly decaying $\mathrm{Z}$ has the detector signature of missing transverse momentum, $p^{\mathrm{miss}}_{\mathrm{T}}$, and reconstructing this quantity with high fidelity is important across a broad range of LHC analyses.  In the $\mathrm{H}(b\bar{b})+$jets process, jet objects -- collimated sprays of many particles -- are produced.  Furthermore, the substructure of the jet is also very important for a wide array of applications and both jets and jet substructure reconstruction can be affected by the presence of pileup particles. When studying the performance of our algorithm, we use the resolution of the reconstruction of $p^{\mathrm{miss}}_{\mathrm{T}}$ and jet $p_\mathrm{T}$ and mass as benchmark metrics. The dataset generation follows a similar setup as in~\cite{Martinez:2018fwc}, with no detector-level effects taken into account.

For each signal process, 30K hard scattering events and 30M pileup QCD events are generated separately using PYTHIA 8.223~\cite{Sjostrand:2014zea} with 4C tune~\cite{Corke:2010yf}. The 30K signal events are then randomly divided into 3 groups for different pileup conditions, each with 10K signal events. Particles from the hard scattering signal process are overlaid at particle level with the ones from pileup events, which are randomly selected with n$_{\text{PU}}$ following a Poisson distribution, centralized at n$_{\text{PU}}$=20, 80, and 140, respectively. The overlaying process is done through DELPHES 3.3.2~\cite{deFavereau:2013fsa}. All particles with $\pt>0.5\,\mathrm{GeV}$ are kept in the output collection. 
\section{Methodology}\label{chp:method}
The details of the problem formulation and the semi-supervised training setup are provided in this section.

\subsection{Formulating pileup mitigation as a graph-based SSL problem}\label{sec:rationale}

Graph-based SSL is a widely-used technique in the ML community to handle the case where training samples (labeled) and testing samples (unlabeled) are connected as nodes in a graph~\cite{zhu2005semi,kipf2017semi}. More importantly, the graph structure that connects these nodes also indicates a certain level of labeling information. For example, it is widely used to detect social-group labels of individuals in social network analysis~\cite{fortunato2010community,lancichinetti2009community,li2019optimizing}, where two individuals denoted by two nodes are more likely to be connected if they share the same social-group label. Graph-based SSL effectively combines the graph structure, the labels of training samples, and the features of both training and testing samples all together to predict the labels of testing samples with high accuracy.

The pileup mitigation problem can be naturally formulated as a graph-based SSL problem by utilizing the geometric relationship between charged particles and neutral particles, and the labels of charged particles to make predictions over neural particles. Specifically, as introduced in Section~\ref{chp:intro} and~\ref{chp:related}, charged particle labels (LV or PU) can be precisely determined in real experiments for the most part, while neutral particle labels remain unknown and need to be inferred. Because of the short range of hadronization and parton showers, and also the larger boosts from higher-\pt energetic particles, charged and neutral particles from the LV tend to be more localized in certain regions of the $\eta-\phi$ space, while particles from PU are more isotropically distributed. Exploration of such local connections between charged and neutral particles helps identify individual particles produced from LV or PU. Therefore, an effective learning procedure of a model should not only leverage the self features of the neutral particle, but also the features of its neighboring particles, in particular the labels of its neighboring charged particles, which the graph-based SSL is by definition designed for. For each event in our study, we view particles as nodes and connect particles with edges if their distance in the $\eta-\phi$ space is small. Note that building graph formulation also naturally fits the sparse nature of the particles located in the geometric space. Alternative ways such as viewing the data as images in the $\eta-\phi$ space with regular pixels and rounding the locations of particles onto those pixels, often suffer from the rounding error and a granularity selection issue.

\paragraph{Unique ML aspects of the pileup mitigation problem} \label{sec:method}

There are also two fundamental differences between the pileup mitigation problem and a traditional graph-based SSL problem to be noted:
\begin{enumerate}
    \item \textbf{Graph-level generalization.} Pileup mitigation requires graph-level generalization that traditional graph-based SSL does not need. Traditional graph-based SSL typically adopts only one single graph, e.g., a social network, to connect all training and testing samples. In pileup mitigation, each event forms one graph consisting of both charged and neutral particles. The obtained model is also expected to be generalized across different types of events (graphs). In our studies, multiple events (graphs) are used to train and test the model.
    
    \item  \textbf{Particle-level label usage.} The way to use labels in pileup mitigation is fundamentally different from that in traditional graph-based SSL. Traditional graph-based SSL typically assumes that labels of training samples are only used to supervise the model training and not used as input features of the model. However, in pileup mitigation, the labels of charged particles to supervise the model training are also needed to feed the model as features, because they provide very informative information for the inference of neighboring particles. If we use the labels of all charged particles both as the input features of the model and to supervise the model training, the obtained model cannot be applied to the inference of neutral particles as neutral particles do not have such labels. To address such a problem, we propose a random masking strategy, where we randomly mask the charged particles to decide whether their labels are used to supervise the model training or as input features. 
    The detailed masking process will be discussed in the following section.
    \end{enumerate}

\subsection{Detailed approach}
\begin{figure}[!htp]
    \centering
    \includegraphics[width=0.99\textwidth]{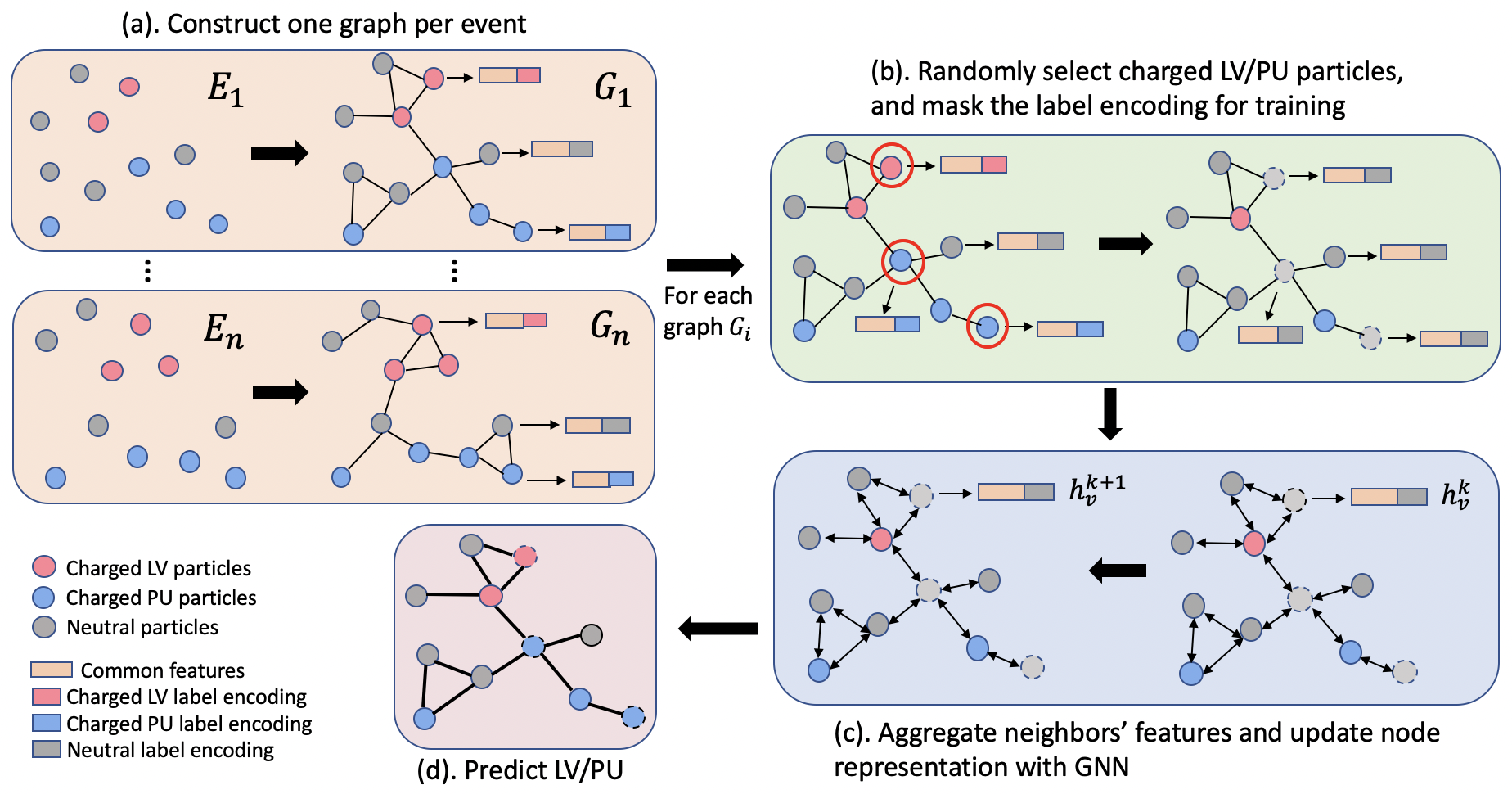}
    \caption{A diagram illustrating the SSL model training flow}
    \label{fig:model_diagram}
\end{figure}

The four steps of the developed approach to train the model are provided in Figure \ref{fig:model_diagram}.
\begin{enumerate}[label=(\alph*)]
    \item Graphs are constructed on an event-by-event basis where each node in the graph is one particle;
    \item A random selection and masking of charged particles are carried out; 
    \item A GNN is applied to aggregate neighboring features and update the node representations;
    \item The LV/PU prediction is computed based on the final node representations. 
\end{enumerate}
Details are documented in the following subsections. It is worth pointing out that although the GNN takes the entire graph as input, only those selected and masked charged particles will be used to supervise the model training, specifically for computing the loss function, performing backward propagation, and optimizing the model parameters. At the inference stage, the masking procedure is excluded and the inference is conducted on all neutral particles. We explain the details of the approach as follows.

\subsubsection{Graph construction}
One graph is constructed per event to establish the relations between particles and their neighbors. Particles are viewed as nodes and two particles are connected if their distance in the $\eta-\phi$ space, $\Delta R=\sqrt{(\Delta\eta)^2 + (\Delta\phi)^2}$, is smaller than a certain threshold $R_0$. A smaller threshold would result in a sparser graph, easier to compute but with less neighbor information, while a larger threshold would result in a denser graph, with more neighbor information but more computing-intensive. $R_0=0.4$ is chosen in this study.

\subsubsection{Graph Neural Networks}
Graph Neural Networks (GNN) have shown to be one powerful tool for graph-based SSL~\cite{kipf2017semi}. GNN first associates each node with a node representation based on the initial node features and then updates node representations by aggregating and combining with the representations of the neighboring nodes. One widely-used GNN model is GraphSage~\cite{hamilton2017inductive}, which takes an average of neighboring node features for node representation update. Several previous works in applying machine learning techniques for pileup mitigation also chose GNN as their models~\cite{Martinez:2018fwc, Maier:2021ymx}.

Even though many varieties of GNN models can be applied to our framework in pileup mitigation, we focus on using a variant of the gated GNN model~\cite{li2016gated}. Since there are certain scenarios where LV particles are surrounded by PU particles, the gated GNN model can automatically learn the gates to control the aggregation of neighboring particles' representations. In contrast, the GraphSage model does not have such control when averaging the representations of the neighbors.

\begin{figure}[t]
    \centering
    \includegraphics[width=0.99\textwidth]{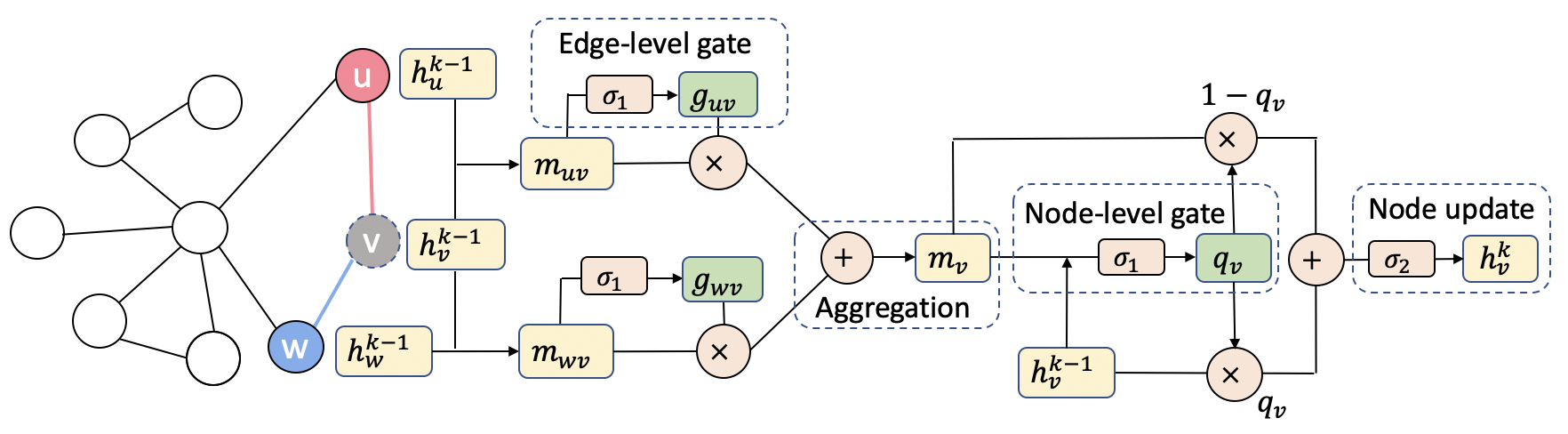}
    \caption{Node representation $h_v$ update for the $k^{th}$ iteration.
    }
\label{fig:gated_model}
\end{figure}

Let $h^{k}_v$ denote the node $v$ representation at $k$-th layer. Our gated GNN model, as shown in Figure~\ref{fig:gated_model}, is formulated as following:
\begin{align}
    \text{Message formulation:}&\quad m_{uv} = \left[h_u^{k - 1}, h_v^{k - 1}, \Delta\eta_{uv}, \Delta\phi_{uv}, \Delta R_{uv}, h_g^{k - 1}\right] \label{1}, \\
    \text{Aggregation:}&\quad  m_v = \textstyle{\sum_{u \in N(v)}}{g_{uv} m_{uv}}, \,\text{where}\,\, g_{uv} = \text{Sigmoid}(W_1  m_{uv} + b_1)\, \label{2}\\
    \text{Node-level gate:}&\quad  q_v = \text{Sigmoid}(W_2  [h_v^{k - 1}, m_v] +b_2) \label{3}\\
    \text{Node update:}&\quad  h_v^k = \text{ReLU}(q_v (W_3 h_v^{k - 1} + b_3)) + (1 - q_v)  ( W_4 m_v + b_4)), \label{4}% \,\text{where}\,\, q_v = \text{Sigmoid}(W_2  [h_v^{k - 1}, m_v, h_g^k] +b_2)
\end{align}
where $\Delta\eta, \Delta\phi, \Delta R$ are the geometric features that characterize two particles' spatial coordinates differences $\eta$, $\phi$, and distance $\Delta R = \sqrt{\Delta\eta^2 + \Delta\phi^2}$, and $h_g$ is a global node, which is calculated as the average of all node representations in one graph. The node representations are initialized as particle features that in our studies include the particle transverse momentum $p_\mathrm{T}$ and one-hot label encoding, that is, $(1,0,0)$ for PU charged particles, $(0,1,0)$ for LV charged particles, $(0,0,1)$ for neutral particles and masked charged particles, where the procedure of masking charged particles will be introduced in Sec.~\ref{sec:mask}. For a target node $v$, in Eq.~\eqref{2}, $g_{uv} \in [0,1]$ is a weight learned for each neighboring node $u\in N(v)$ to control the amount of information that is passed to $v$. In Eq.~\eqref{3}, another gate $q_v \in [0,1]$ controls the portion between the representation at $(k-1)$-th layer $h_v^{k-1}$ and the aggregation from the neighbors $m_v$, when formulating the new node representation $h_v^k$ in Eq.~\eqref{4}. The node representations of the selected particles in the final layer of the GNN are put through a multi-layer perceptron~\cite{LeCun2015} with two hidden layers to make the final prediction.

\subsubsection{Masking charged particle and random selection} \label{sec:mask}
The primary goal of masking a subset of charged particles is to make the model leverage the labels (LV or PU) of charged particles in two different ways simultaneously. On the one hand, the masked charged particles are used to supervise the model training, with the expectation that the model trained on these charged particles can be applied to infer the labels of neutral particles in the later testing stage. Therefore, the features of these charged particles in the training should mimic the ones of neutral particles, and their LV/PU labels should not be used as the input features. On the other hand, the LV/PU label information of neighboring charged particles serves as important inputs for predicting the labels of target particles. Thus the label information of neighboring particles should be kept in the inputs.

Note that such masking procedure is at risk of breaking the original structure of the data and thus may introduce biases. To reduce such bias, our model only masks a small portion of the charged particles per event. However, masking only a small portion of charged particles for training may not sufficiently leverage the labels. To achieve a better usage of the labels, we propose the \emph{random selection mechanism}. That is, for each event, we perform multiple-time random selections of the charged particles for masking. This guarantees that for each event each time, only a small portion of charged particles are masked and used to supervise the model training, while most of the charged particles of this event can be eventually used to supervise the model after running the model multiple times on this event. 

Another practical consideration is regarding the time complexity of model training. Although the random masking strategy guarantees a sufficient usage of the labels and the data, setting the masking portion too small may slow down the training procedure, because a huge time of model running per event is needed to guarantee a good coverage of the masking procedure.  
To balance the tradeoff, in our experiments we randomly select about 10$\%$ of charged particles per event each time. Table~\ref{tab:nparticlesselect} includes the numbers of selected charged LV and PU particles per graph per epoch and the total number of charged LV and PU particles per graph. With about tens of training epochs, all charged particles should be selected as training data at least once by random selection. Even though different pileup levels seem to affect the actual numbers of selected particles greatly, experiments show that the model is robust when it is trained on one pileup level and tested on another pileup level.

\begin{table}[t]
    \centering
    \small
    \begin{tabular}{c|c|c|c|c|c|c}
        \hline\hline
        \multirow{2}{*}{$\mathrm{n}_\text{PU}$} & \multicolumn{4}{c|}{\# Particles (in total)} & \multicolumn{2}{c}{\# Selected Particles (for training)} \\
        \cline{2-7} 
        & Charged LV & Charged PU & Neutral LV & Neutral PU & Charged LV & Charged PU\\\hline
       20 & $69\pm34$ & $407\pm154$ & $43\pm22$ & $203\pm80$ & 7 & 35\\\hline
       80 & $67\pm33$ & $1630\pm310$ & $42\pm22$ & $813\pm160$ & 7 & 126\\\hline
       140 & $68\pm34$ & $2846\pm408$ & $43\pm22$ & $1423\pm213$ & 7 & 224\\
       \hline\hline
    \end{tabular}
    \medskip
    \caption{The first four columns include the average numbers of four types of particles under different pileup conditions per graph. The last two columns indicate the number of charged particles being randomly selected for training per graph in one epoch.}
    \vspace{-4mm}
    \label{tab:nparticlesselect}
\end{table}

\subsection{Training details and complexity}
We also compare our SSL model with a fully supervised model that has the same architecture but is trained using the labels of neutral particles without any masking strategy. The fully supervised model needs to be trained and tested over different events, though our SSL model does not need to. To make fair comparisons, for experiments where $\mathrm{n}_\text{PU}=80$, there are 3000/1000/1000 events for training/validation/testing. When $\mathrm{n}_\text{PU}=140$, 2000/800/800 events are used for training/validation/testing. For the $\mathrm{n}_\text{PU}=140$ scenario, there are more particles per event, so the total number of events is reduced to maintain reasonable memory usage. To save graph construction time for random masking, a masking vector is implemented to efficiently mask the charged particles for training each epoch. The vector can be easily altered to mask another set of training particles without constructing a new graph entirely. During training, the model is trained until convergence, which normally takes about 5 times running over all the events for training. The total number of parameters is around 1300 and can be trained within 6 hours on one NVIDIA Tesla V100 or P100. 

In order to reduce the training complexity, we construct graphs by only connecting particles with $\Delta R \leq 0.4$. This restriction in $\Delta R$, in this case, will make the entire graph sparse to reduce the time for graph construction, training, and inference. The graph construction time is approximately 0.1\,s per event (per graph) and the inference time is about 30\,ms for a graph with $\Delta R \leq 0.4$ and about 50\,ms for a graph with $\Delta R \leq 0.8$ at $\mathrm{n}_\text{PU}=80$. The inference time becomes longer if we increase $\Delta R$ when constructing the graph.

\section{Results}\label{chp:results}
Experiments are carried out to verify the effectiveness of the model trained via SSL and its ability to be adapted to different $\mathrm{n}_\text{PU}$ levels. The performance with PUPPI, semi-supervised training, and supervised training are compared in this section. Firstly, we examine the performance at particle level, using the receiver operating characteristic (ROC) curves and the area under the ROC curve (AUC) scores trained and tested under different pileup conditions. Then, the performance of physics observables, such as the hadronic jet mass and \pt, and also the missing transverse momentum $p^{\mathrm{miss}}_{\mathrm{T}}$, are studied and compared among the three approaches. Finally, some event display examples are provided, visualizing the differences and improvements of (semi-)supervised results with respect to PUPPI.

\subsection{Performance at particle level}
\begin{figure}[!ht]
\centering
\includegraphics[trim={1.5cm 3.7cm 12.7cm 4.08cm},clip,width=0.445\textwidth]{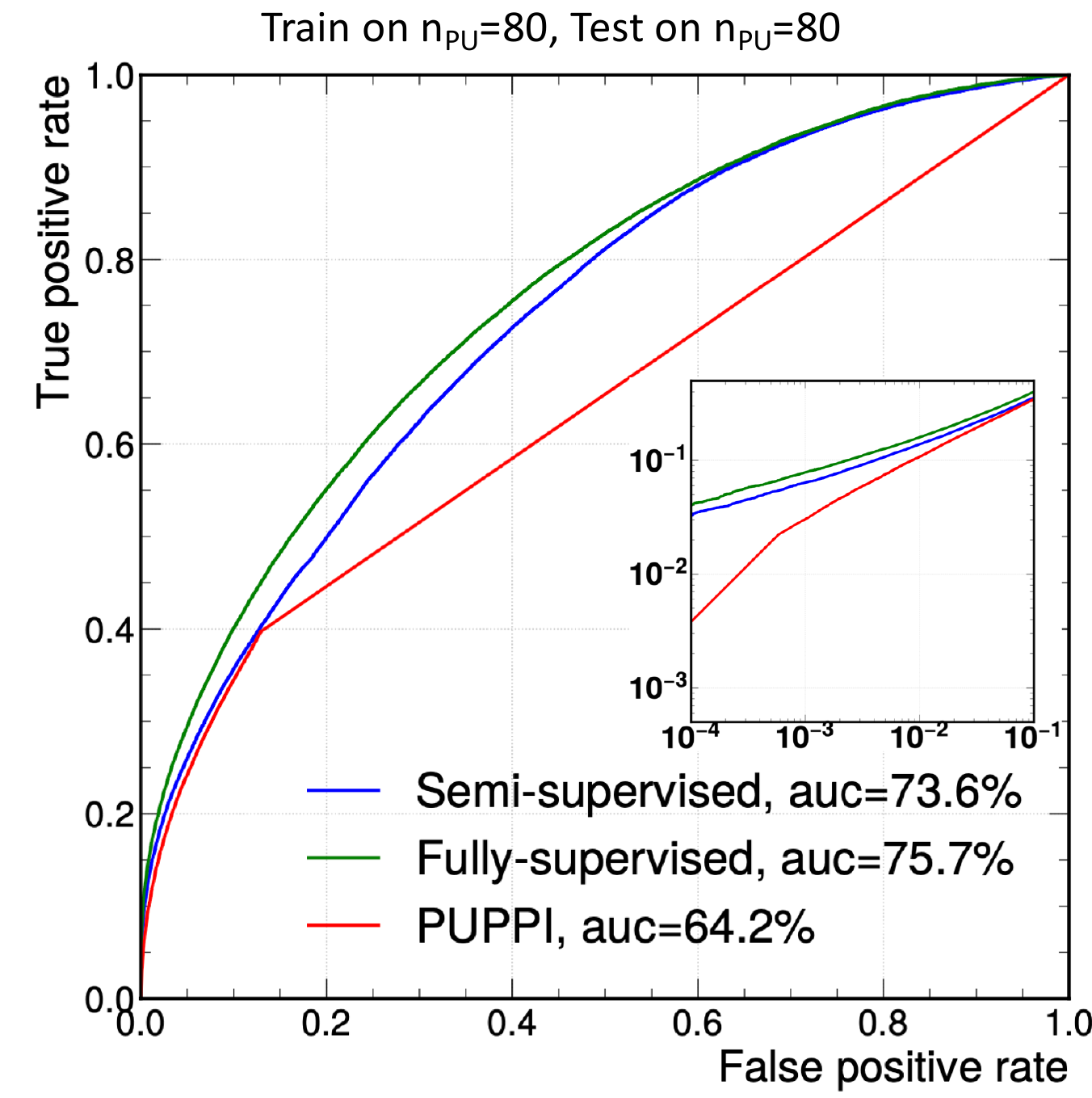}
\includegraphics[trim={1.5cm 3.7cm 13.0cm 4.08cm},clip,width=0.445\textwidth]{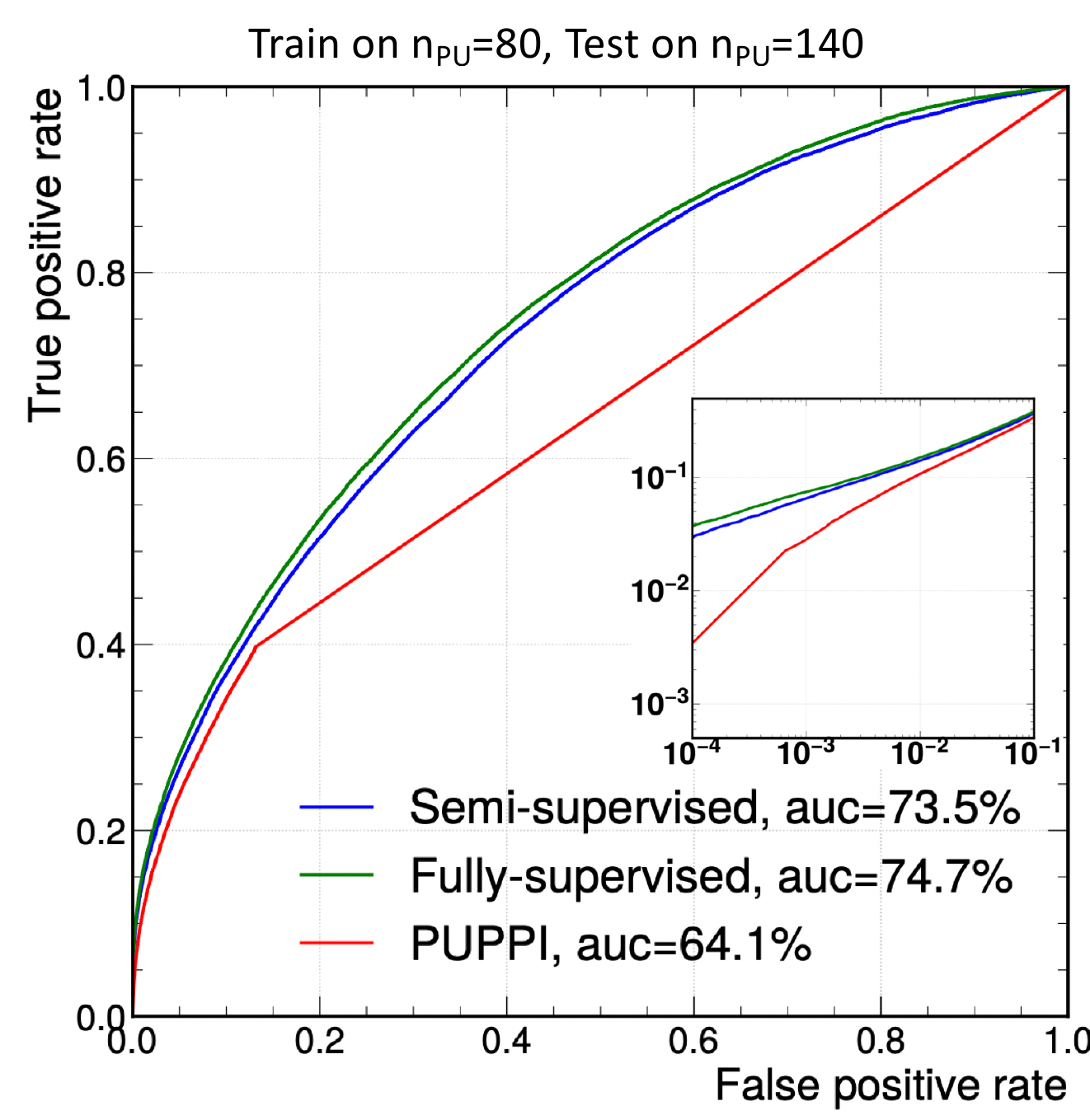}
\includegraphics[trim={1.5cm 3.7cm 12.9cm 4.08cm},clip,width=0.445\textwidth]{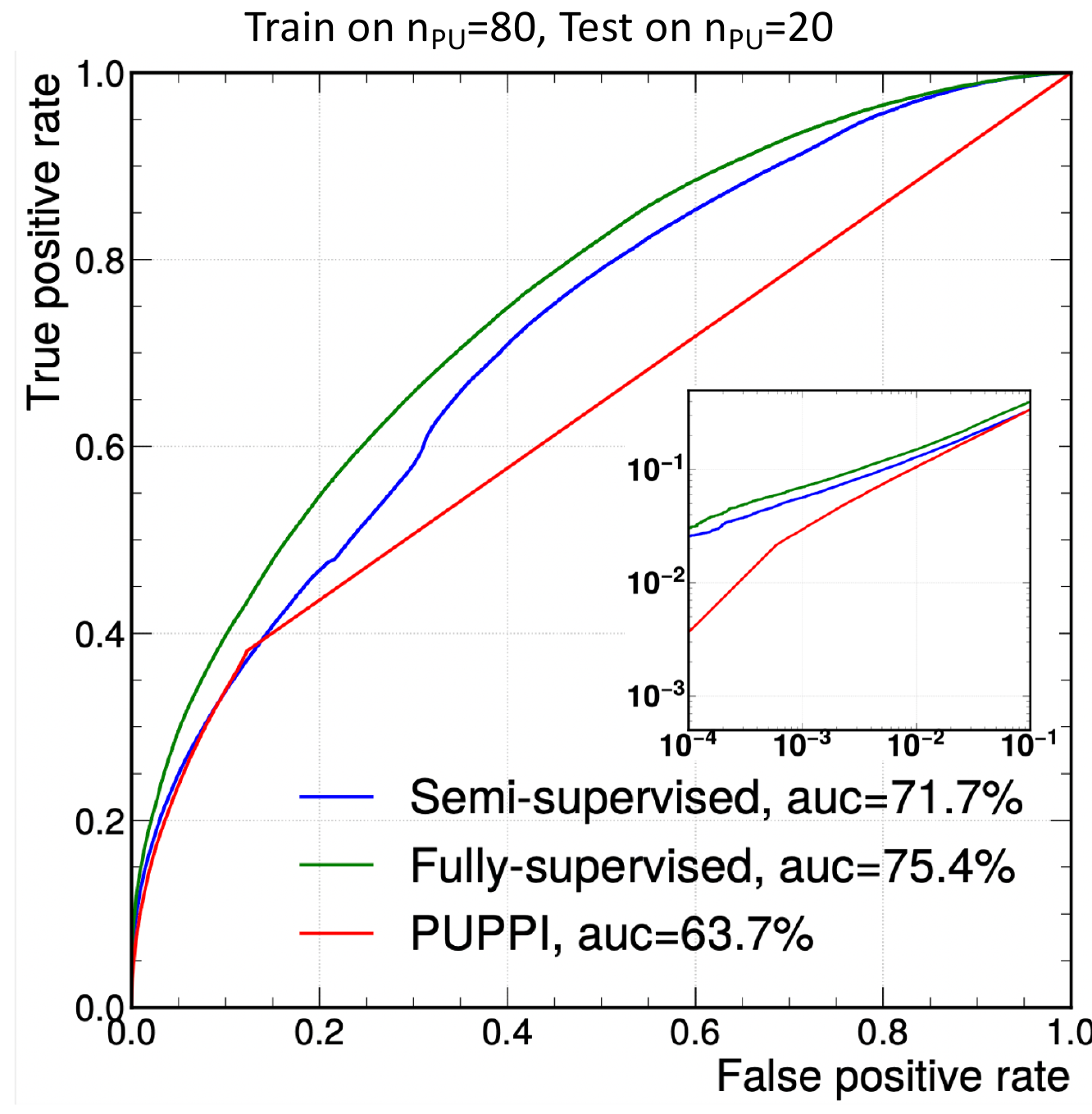}
\caption{The ROC curves for the gated GNN on neutral particles under $\mathrm{n}_\text{PU}=80$ for SSL, fully-supervised learning, and the domain PUPPI algorithm. The small plot inserted under the ROC curve is the log scale of the lower left region of the ROC curve for better visualization.}
\label{fig:ROC}
\end{figure}

Figure~\ref{fig:ROC} shows the ROC curves and Table~\ref{table:combined_scores} lists the area under the ROC curve (AUC) score trained and tested in different $\mathrm{n}_\mathrm{PU}$ conditions.  These plots show the performance at a per-particle level of the LV and PU labels. In all these cases, when training and testing at the same pileup level, both SL and SSL outperform PUPPI by around 10\%, and the performance decrease from SL to SSL is within a few percent. An inset shows the plot in the log-log scale where smaller false positive rates are important given the much larger number of PU particles compared with the number of LV particles in one event. How the per-particle performance is manifested in physics object performance will be explored below.  

When training and testing under different pileup conditions, SL models seem to be more robust to different pileup conditions, i.e., the AUC scores are more consistent across different training and testing conditions, whereas the SSL model is more sensitive to the pileup condition. An interesting observation is that the SSL model performs well when extrapolating to a higher $\mathrm{n}_\text{PU}$ condition, but not the reverse order. It is interesting to further improve the generalization capability of the model across different pileup conditions. We leave it as a future research direction.

\begin{table}[ht]
\centering
\begin{tabular}{ c|c|c|c|c} 
\hline\hline
\diagbox[]{Training}{Testing} & Model type & {$\mathrm{n}_\text{PU}=20$} & {$\mathrm{n}_\text{PU}=80$} & {$\mathrm{n}_\text{PU}=140$} \\
\hline
\multirow{3}{*}{$\mathrm{n}_\text{PU}=20$}& PUPPI & 63.6 & 64.2 & 64.1\\
\cline{2-5} 
& SL & 76.6 & 75.5 & 74.3\\ 
\cline{2-5} 
& SSL & 73.9 & 70.7 & 67.6\\ 
\hline

\multirow{3}{*}{$\mathrm{n}_\text{PU}=80$}& PUPPI & 63.7 & 64.2 & 64.1\\
\cline{2-5} 
& SL & 75.4 & 75.7 & 74.7\\ 
\cline{2-5} 
& SSL & 71.7 & 73.6 & 73.5\\ 
\hline

\multirow{3}{*}{$\mathrm{n}_\text{PU}=140$}& PUPPI & 63.7 & 64.2 & 64.1\\
\cline{2-5} 
& SL & 69.5 & 75.2 & 75.0\\ 
\cline{2-5} 
& SSL & 53.0 & 71.7 & 72.9\\ 
\hline\hline

\end{tabular}
\caption{AUC scores (\%) of the PUPPI algorithm, supervised (SL) and semi-supervised (SSL) models trained and tested on three different pileup conditions: $\mathrm{n}_\text{PU}=$20, 80, and 140. }
\label{table:combined_scores}
\end{table}

\subsection{Performance on jet observables}
The GNN model output, which is an N-dimensional array of float numbers between 0 and 1 (N is the total number of particles per event), can be interpreted as the probability of how likely each corresponding particle is produced from the LV. Similar to the approach adopted in the PUPPI algorithm, the four-momenta of all particles are rescaled with the corresponding GNN outputs. The jets are then clustered with the particle rescaled four-momenta using the anti-kt jet reconstruction algorithm~\cite{Cacciari:2008gp}, with $\Delta R$ chosen to be 0.7 to be consistent with previous related work~\cite{Martinez:2018fwc}. Jets clustered with the generator-level LV particles serve as the ground truth information for comparison.

\begin{figure}[!htp]
    \centering
    \includegraphics[width=0.45\textwidth]{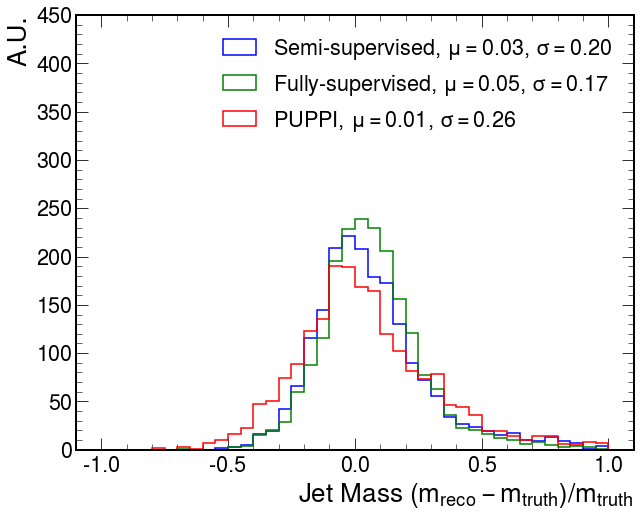}
    \includegraphics[width=0.45\textwidth]{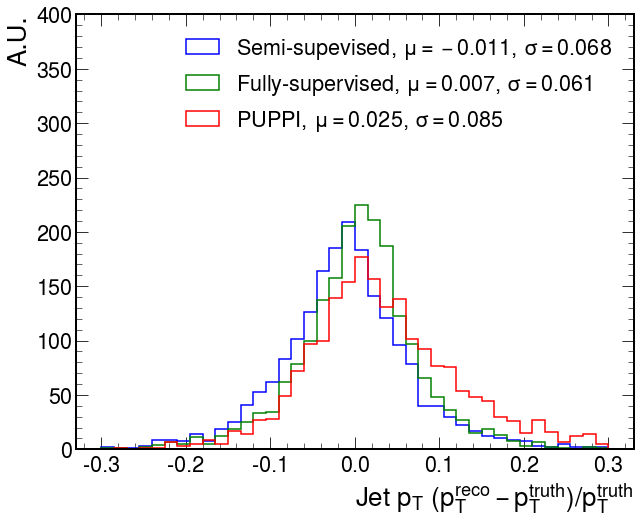}
    \caption{Performance on jet mass and jet \pt with different pileup mitigation techniques for $\mathrm{n}_\text{PU} = 80$.}
    \label{fig:jet_mass}
\end{figure}

We study the leading jet in the event with truth \pt above 20\,GeV in the $\mathrm{H}(b\bar{b})+$jets sample. Because this process is inclusive, the typical jet \pt is approximately 60\,GeV, nearly half the Higgs mass. Figure~\ref{fig:jet_mass} shows the reconstructed jet mass and \pt resolutions with respect to the truth-level jets for the scenario where $\mathrm{n}_\text{PU} = 80$.  Resolutions are studied for reconstructed jets which are within $\Delta R = 0.1$ of a truth-level jet. Compared with PUPPI, the bias and resolution of jet masses and \pt clustered with both the semi-supervised and fully-supervised algorithms are significantly smaller. This indicates that the GNN approach does both a better job in predicting the overall aggregate \pt of the jet object, and also its substructure using the jet mass metric.  Improvements over PUPPI are comparable to other DL approaches using GNNs~\cite{Martinez:2018fwc}.  Compared with the SL approach, the performance drop of the SSL approach is relatively small. These are consistent with the per-particle performance results and show the improvements provided by the SL and SSL models.  The results are also consistent across different pileup scenarios.

\subsection{Performance on missing transverse momentum}
\begin{figure}[!htp]
    \centering
    \includegraphics[width=0.48\textwidth]{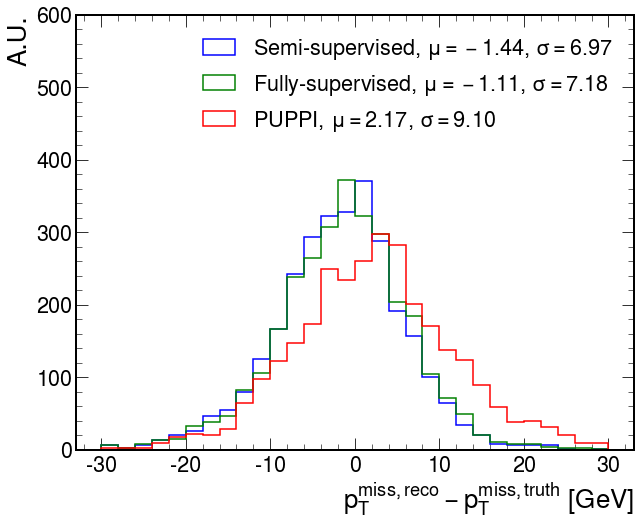}
    \caption{Resolution of the missing transverse momentum of events for the different pileup mitigation models for $\mathrm{n}_\text{PU} = 80$.}
    \label{fig:met_pt}
\end{figure}

We study the missing transverse momentum ($p^{\mathrm{miss}}_{\mathrm{T}}$) resolution performance of our algorithm using $\mathrm{Z}(\nu\nu)+$jets events.  The $p^{\mathrm{miss}}_{\mathrm{T}}$ is the negative vector sum of the particles in the event and are calculated with the rescaled four-momenta of all particles. We compare the SL and SSL approaches with the performance of PUPPI and the results are shown in Figure~\ref{fig:met_pt}. Compared with PUPPI, the resolution is significantly better ($\sim 20\%$), with some minor deviation in the mean value from zero. This can potentially be due to the SSL misidentifying some LV particles as PU ones, and removing these from the LV collection. The bias nevertheless is small and can be mitigated via offline calibrations. The results are also consistent across different pileup scenarios.

\subsection{Event visualization}
\begin{figure}[!htp]
    \centering
    \includegraphics[width=0.45\textwidth]{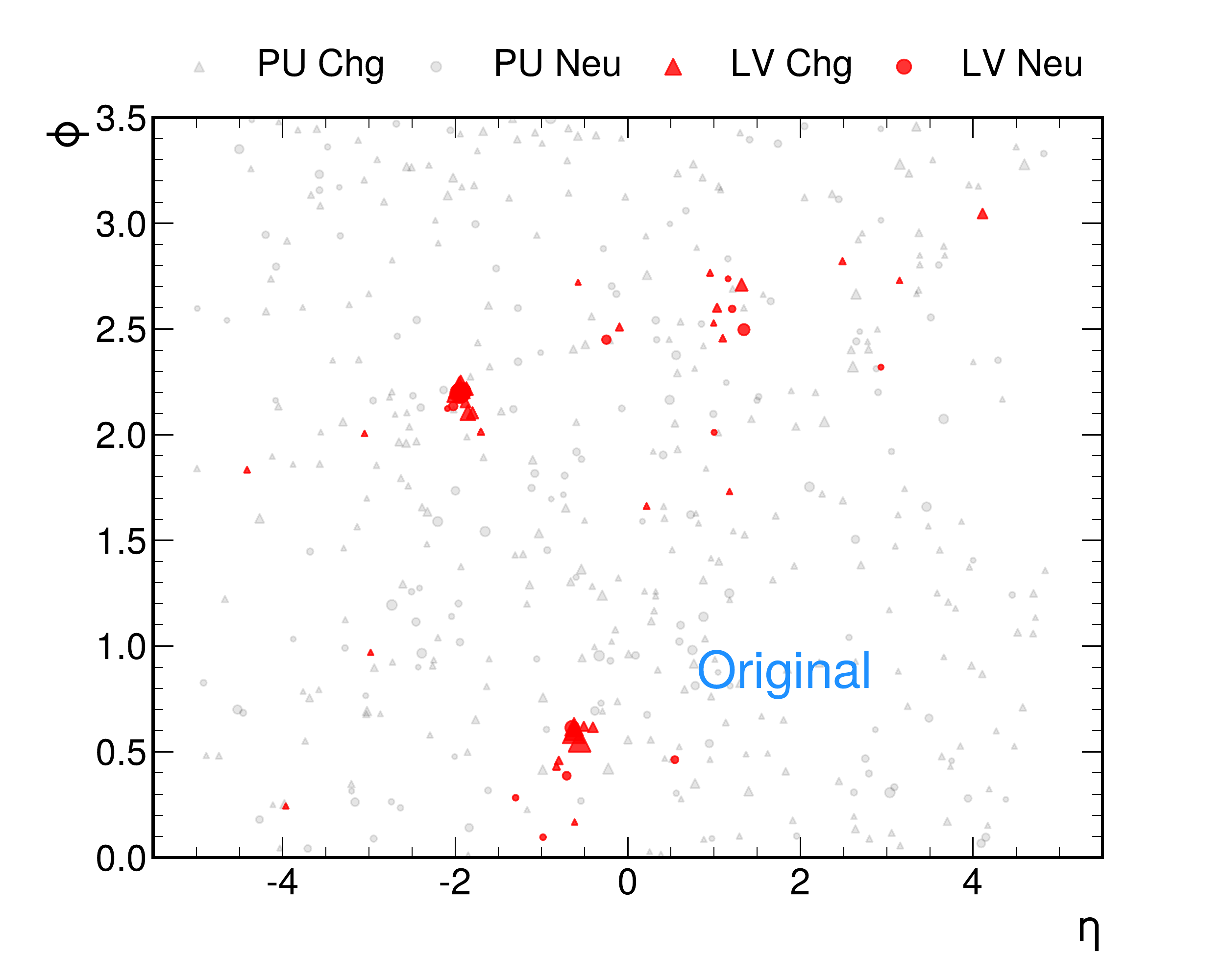}
    \includegraphics[width=0.45\textwidth]{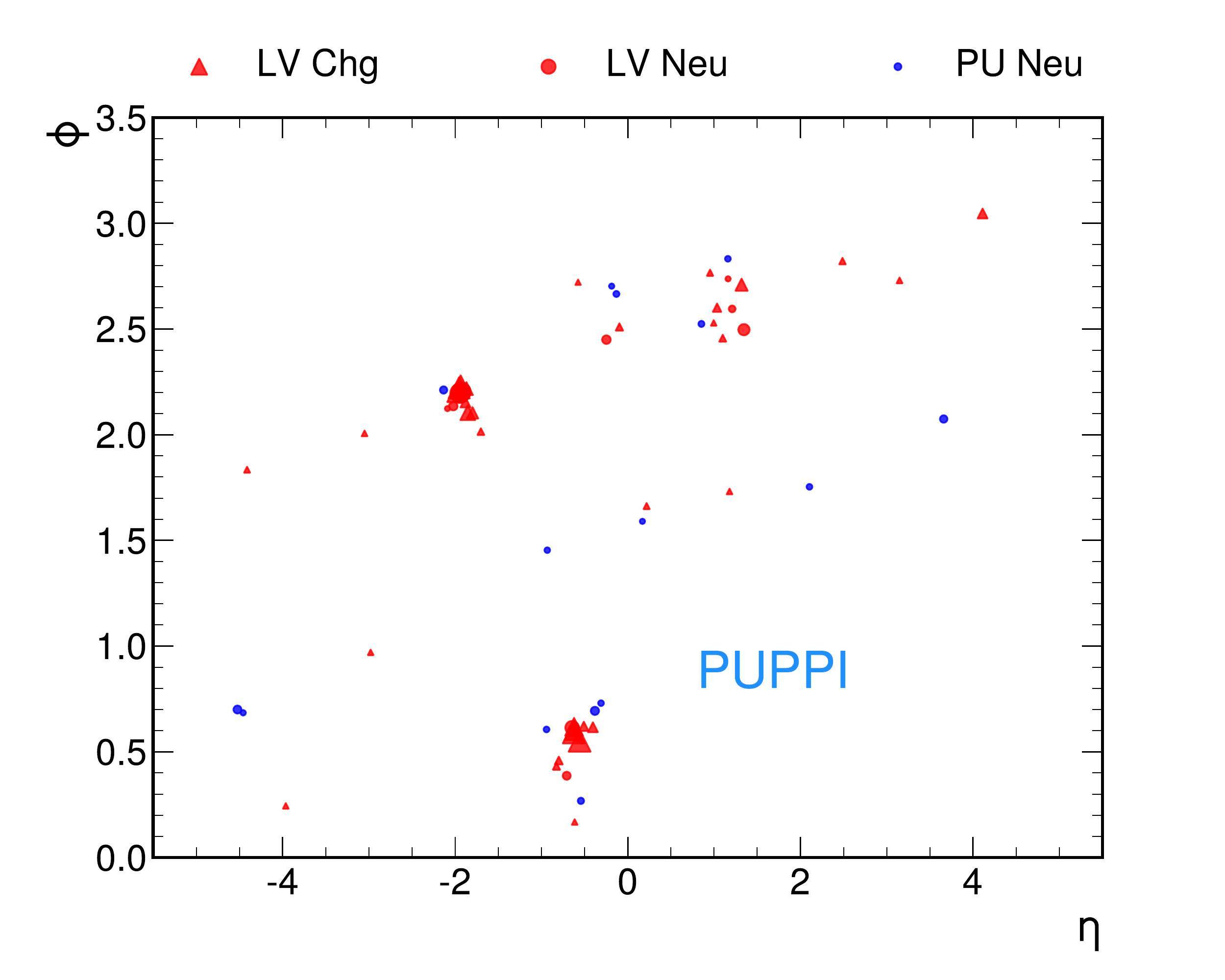}
    \includegraphics[width=0.45\textwidth]{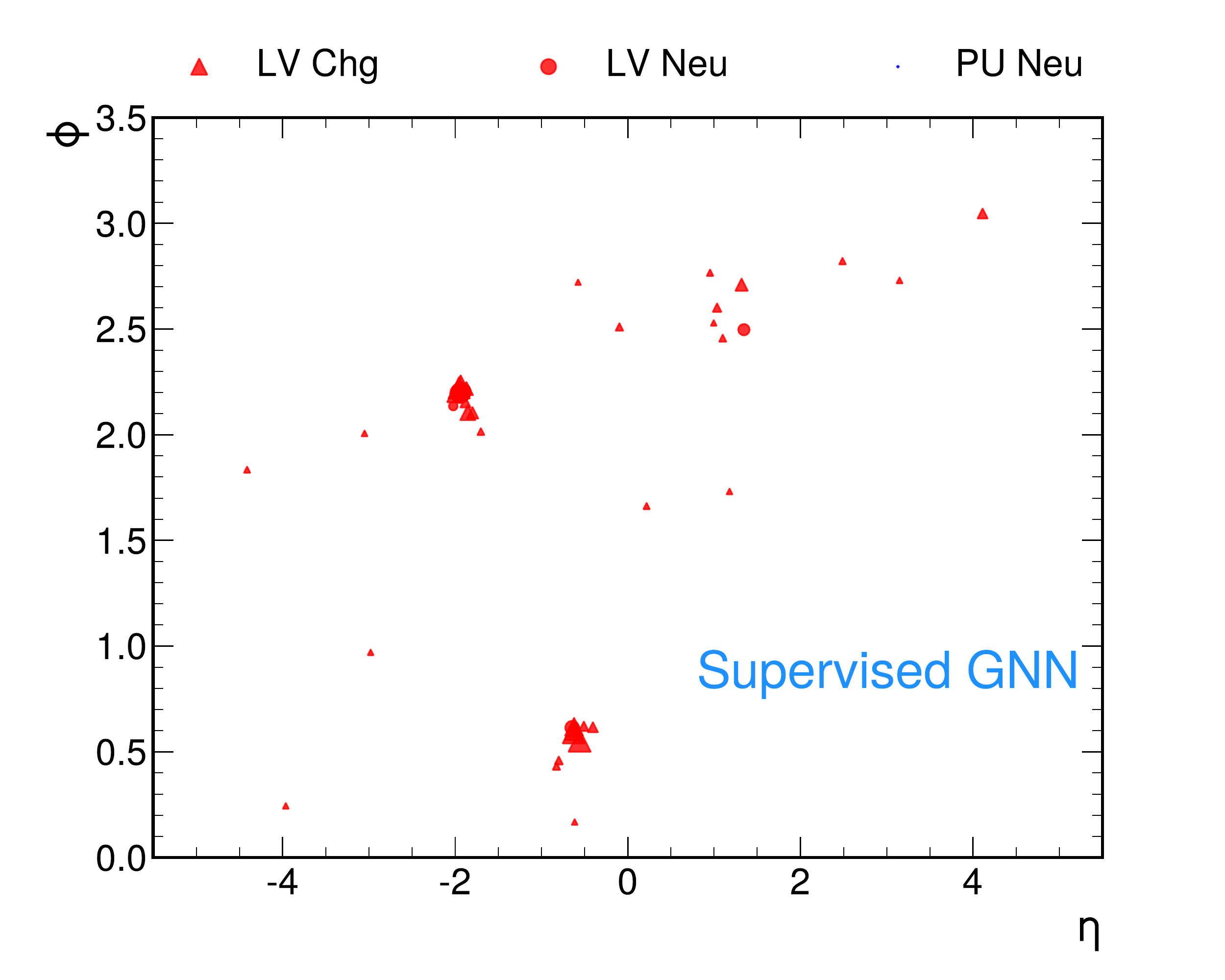}
    \includegraphics[width=0.45\textwidth]{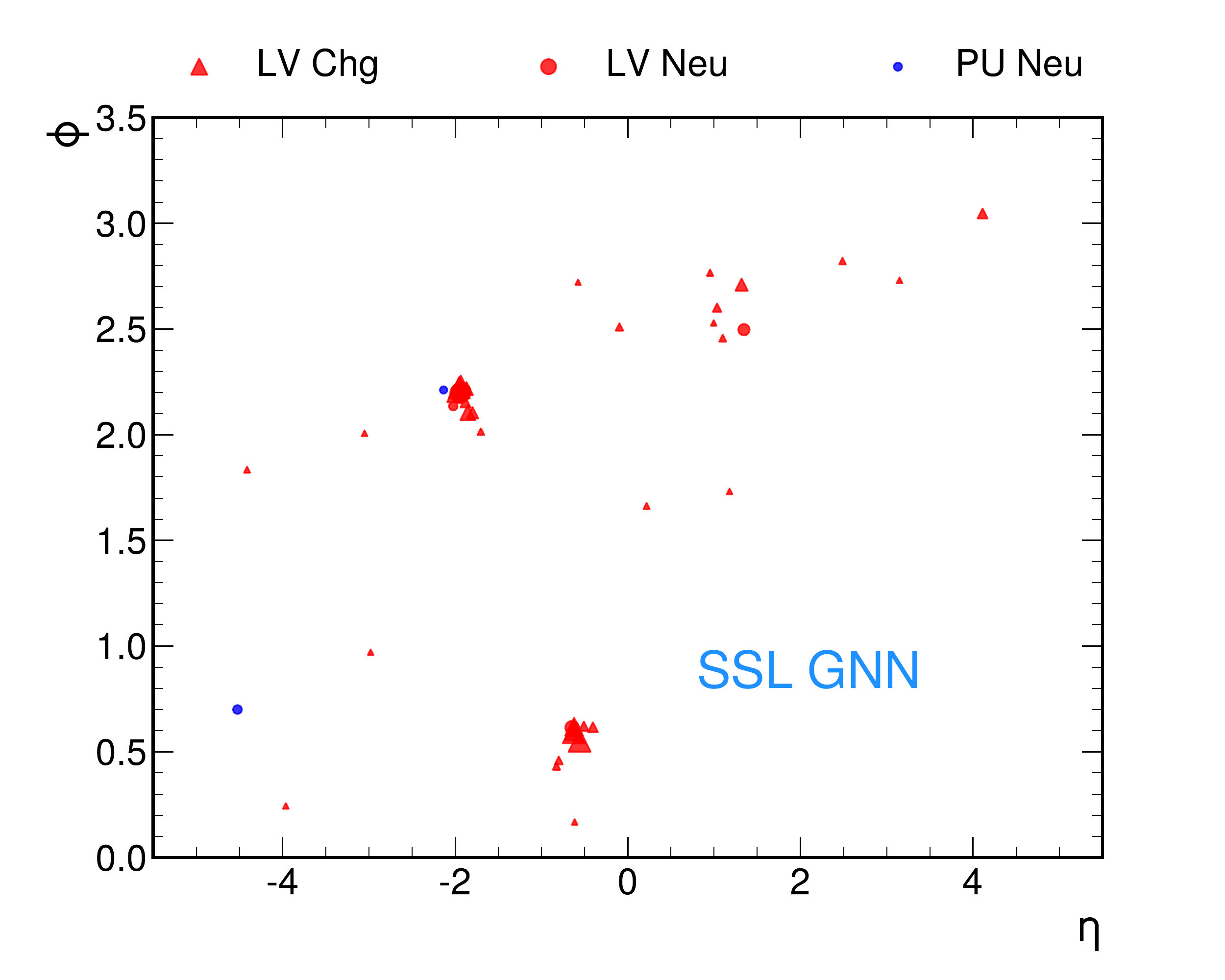}
    \caption{Some event display examples of the particle distributions in the $\eta-\phi$ space. The upper left plot is using ground truth information. The upper right plot is after applying PUPPI. The lower left and right plots are after applying SL and SSL, respectively.}
    \label{fig:evtdisplay}
\end{figure}

Figure~\ref{fig:evtdisplay} provides one event visualization of the particle distributions in the $\eta-\phi$ space with different pileup mitigation algorithms: PUPPI (top right), SL (lower left), and SSL (lower right). The marker size scales with the particle \pt. It can be observed that while PUPPI leaves some PU remnants, both SL and SSL models clean the PU particles more efficiently while preserving the LV particles.

In summary, the performances of all these particle-level metrics and physical observables are consistent and show the improvements of the SSL models with respect to the currently widely used domain algorithm PUPPI. Compared with the traditional SL approach, the performance decreases with the novel SSL approach is negligible. However, the SL approach cannot be directly applied to the real experimental data while the SSL approach can. The trainings and evaluations are currently all performed on the {\tt DELPHES}-based simulation data, and their effectiveness will be carefully re-examined on the {\tt GEANT}-based simulation and real collision data in future studies.

\section{Discussion on the connections with PUPPI}\label{chp:discussion}
This section briefly discusses the connections between the Graph (S)SL model and PUPPI, in both the algorithm design and outputs, in order to help provide some insights into what the model learns and where further improvements could be realized.

\subsection{Model design and understanding behavior}
As briefly mentioned in previous sections, the GNN model architecture is designed to mimic the $\alpha$ calculation in PUPPI, with trainable parameters directly learned from data that can be more expressive and powerful. The similarities and differences between PUPPI and the GNN model are compared in detail here:
\begin{enumerate}
    \item \emph{Targets particle self features}. PUPPI does not use particle self features while the GNN model does. In PUPPI, only the neighboring particle features are included when determining PUPPI weights, whereas in the GNN model both the target and the neighboring particle information are used in Eq.~\eqref{4}. This is potentially very useful in some practical cases: for example, for high-\pt particles which are highly likely to be produced from the LV, PUPPI usually requires one additional step to manually assign high weights but the GNN is expected to handle these well automatically. 
    \item \emph{Selections/Gates to remove noise}. When aggregating information from neighboring particles, different selection criteria can be applied to remove noisy information and keep only the useful ones. Within the tracker acceptance, PUPPI uses all the neighboring charged LV particles. In the GNN model, Eq.~\eqref{2} does a similar job - the gate $g_{\mu\nu}$ is applied to determine the weight (importance) of the neighboring information, and therefore the noisy information can be reduced.
    \item \emph{Choice of metric}. For the neighboring particles passing the selection, PUPPI utilizes their \pt and the $\Delta R$ distance with respect to the target particle, and defines the metric $p_{T,j}/\Delta R_{ij}$. Different metric options, such as $(p_{T,j}/\Delta R_{ij})^2$, or $p^2_{T,j}/\Delta R_{ij}$ were also studied in the PUPPI developments. Claiming what is the best choice is ad-hoc and takes a lot of human labor. This is avoided in the GNN model as more information is the neighboring particles are included in the inputs, such as the $p_{T}$, $\eta$, $\Delta\eta$, $\Delta\phi$, $\Delta R$, and the metric with more complicated and powerful forms can be learned inside the GNN model.
    \item \emph{Generalization of PUPPI}. The Graph SSL model can also be viewed as a direct generalization of PUPPI because both of them learn or tune their parameters by only using charged particles whose labels are available in real collision data. In contrast, the Graph SL model needs extra labeling information from neutral particles. 
\end{enumerate}

\begin{figure}[htp]
    \centering
    \includegraphics[width=0.45\textwidth]{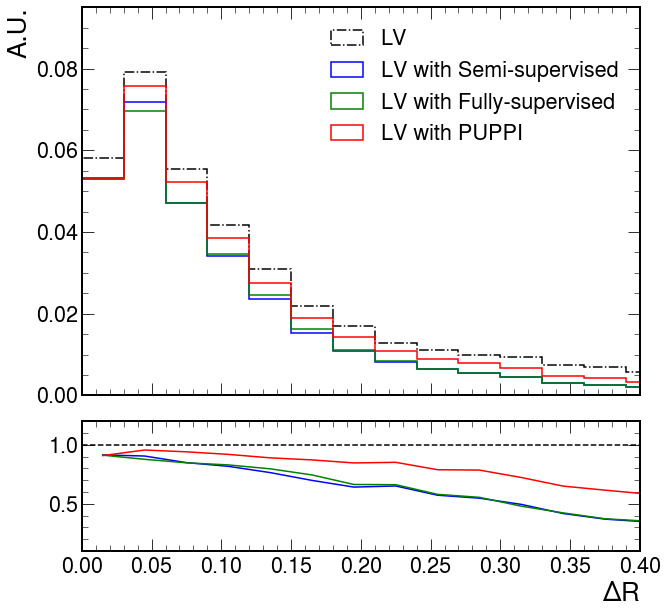}
    \includegraphics[width=0.45\textwidth]{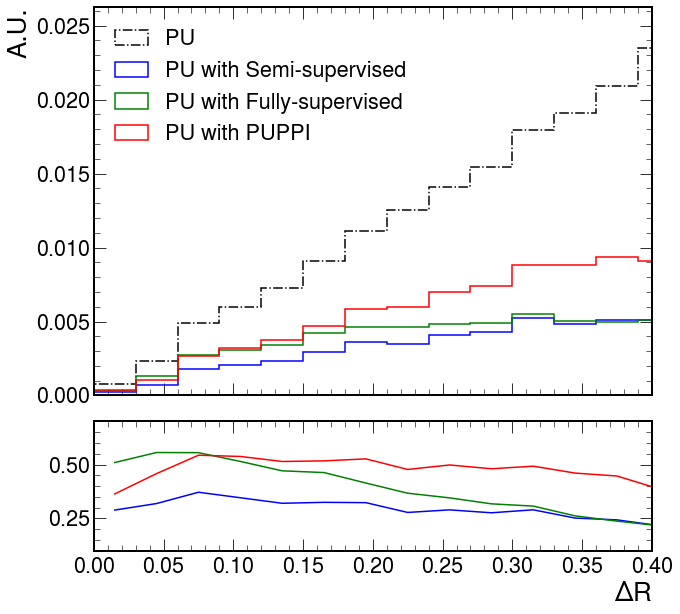}
    \caption{The $\pt$-weighted $\Delta R$ distributions of the LV (left) and PU (right) particles locally to a truth-level jet.}
    \label{fig:deltaR_LVPU}
\end{figure}

Figure~\ref{fig:deltaR_LVPU} shows the $\pt$-weighted $\Delta R$ distribution of the LV (left) and PU (right) particles in the proximity of one truth-level jet in $\mathrm{H}(b\bar{b})+$jets events. The $\Delta R$ is calculated between the particle direction and the associated jet axis, and the particle \pt is normalized to the truth-level jet \pt and served as the weight for each entry in the two histograms.

From the right plot, it can be observed that PUPPI in general removes around 50\% of the PU particles, while the SSL model removes around 75\% of the PU particles. From the left plot, compared with PUPPI, in the central region (small $\Delta R$) where most of the LV particles exist, SL and SSL models keep similar amounts of LV particles as PUPPI. In the region far from the jet axis (large $\Delta R$) where fewer LV particles exist, both SL and SSL models remove more LV particles than PUPPI. From both these plots, it is clear that generally the Graph SL and SSL models are more aggressive in removing particles at the edge of the jets than PUPPI which leads to improved physics performance. However, this also indicates areas of potential further improvements.

\subsection{Output comparison}

\begin{figure}[!htp]
    \centering
    \includegraphics[width=0.98\textwidth]{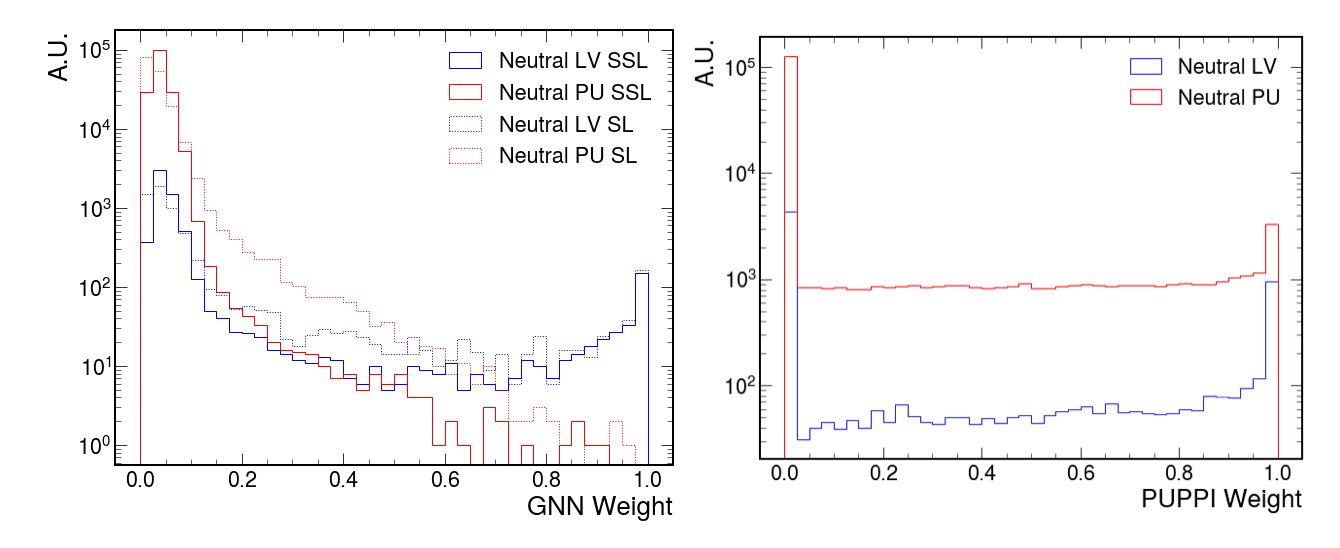}
    \caption{GNN weight (left) and PUPPI weights (right) on neutral particles.}
    \label{fig:gnnweight}
\end{figure}

We would also like to further explore the model outputs and directly compare them to PUPPI outputs.  Figure~\ref{fig:gnnweight} shows the outputs of the GNN model (left) and PUPPI weights (right) for neutral particles. For the pileup neutral particles, most of them get a score close to 0. For the LV neutral particles, a fraction of them get assigned a score close to 1, correctly identified as LV particles, while there are still some assigned a weight close to zero, indicating spaces for future improvements.  In general, the Graph SL and SSL models tend to create a more gradual assignment of LV-like vs PU-like where the weights are bunched more towards 0 or 1 whereas PUPPI will say either definitely the particle is PU or else gives a much more uniform probability. The effect is exacerbated in the case of SL vs. SSL where the SL model tends to give some particles a weight closer to 0.5.  However, given that the plots are presented with a log y-axis scale, these are generally a small fraction of the overall particles.  

\begin{figure}[!htp]
    \centering
    \includegraphics[width=0.98\textwidth]{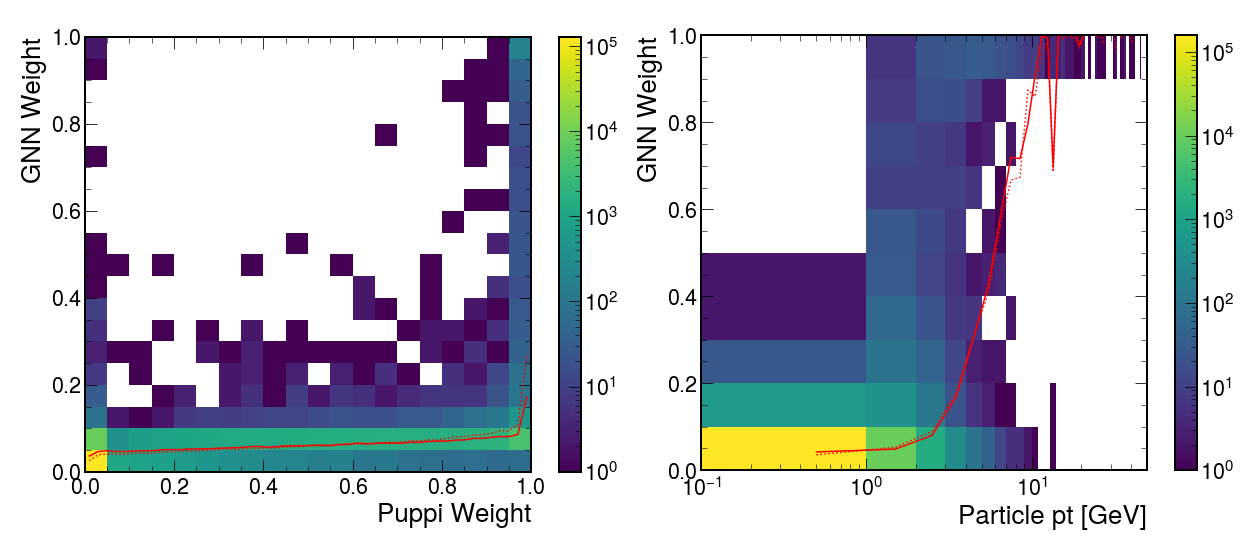}
    \caption{GNN output with respect to the PUPPI weights (left) and the \pt of neutral particles (right).}
    \label{fig:gnnweight_vs_pt}
\end{figure}

Figure~\ref{fig:gnnweight_vs_pt} shows the GNN model outputs with respect to the PUPPI weights (left) and the neutral particle \pt (right). The correlation between GNN outputs and PUPPI weights is not strong, where most of the particles with high PUPPI weights still get relatively small GNN outputs. On the right plot, for particles with \pt below 1\GeV, most of them get assigned weight close to 0; as \pt goes higher, the weight increases, more likely to be produced from the LV.

In summary, it can be observed that the GNN models tend to be more powerful at exploring the high-dimensional input feature space and more efficient in removing pileup noises, especially in the regions with less LV activities (i.e., large $\Delta R$ with respect to the LV jet axes). While the LV particles with high \pt or close to the jet axes are preserved by the GNN model, it is less efficient for keeping the LV information far from the jet axes. This can be further studied and potentially improved in future studies.
\section{Summary and outlook}\label{chp:summary}
This paper presents the first study of semi-supervised ML techniques with a graph neural network for the pileup mitigation task. The task is cast naturally as a graph learning problem, the training is performed on labeled charged particles, and the inference is evaluated on unlabeled neutral particles. This is performed through a careful feature masking process which trains on charged particles \textit{as if} they were neutral particles. By approaching pileup mitigation as a semi-supervised learning problem, we can train from the data and avoid complicated issues arising from (a) data and simulation differences for soft and hard-to-model physics and (b) labeling neutral particles which is inherently challenging given the relatively poor spatial and energy resolution from detecting neutral particles. Compared with PUPPI, the Graph SSL algorithm is more powerful at removing pileup particles, while maintaining the leading vertex particle information. Improvements are observed at the particle-level LV/PU identification and physics observables such as jet \pt and mass, and $p^{\mathrm{miss}}_{\mathrm{T}}$. 

This study serves as a proof of concept, with promising and extensive future studies planned to apply this technique to train directly on real collision data, without any dependence on the ground-truth labeling information. In such cases where the forward region has no tracking information, the momentum and spatial resolutions are expected to be worse than the central ones. We believe that transfer learning techniques can be explored to properly apply the training in the central region to the forward region, and to mitigate the potential larger differences between charged and neutral particles in more realistic scenarios. We show that treating the pileup mitigation task as one that can be machine-learned from data with minimal dependence on simulation is particularly promising and opens up a number of new and interesting challenges for research.  
\section*{Acknowledgements}
We would like to thank Maurizio Pierini and Jean-Roch Vlimant for providing us with the PUPPIML datasets to get started with the training. YF and NT are supported by Fermi Research Alliance, LLC under Contract No.\,DE-AC02-07CH11359 with the Department of Energy (DOE), Office of Science, Office of High Energy Physics and the DOE Early Career Research Program under Award No.\,DE-0000247070. GP and ML are supported by the DOE, Office of Science, Office of High Energy Physics Research Program under Award No.\, DE-SC0007884, and the National Science Foundation (NSF) under award number 2117997 (A3D3). TL, SL, and PL are supported by the NSF award HDR-2117997.

% bibliography
% \newpage
\bibliographystyle{JHEP}
\bibliography{mybib}

\end{document}